\documentclass[usegraphicx,useAMS,usenatbib]{mn2e}

\usepackage{epsf}

\title[The unusual neutron-star transient SWIFT J1749.4--2807]{GRB060602B = 
Swift J1749.4--2807: an unusual transiently accreting neutron-star
X-ray binary}

\author[R.Wijnands et al.]
       {R. Wijnands$^1$\thanks{E-mail:r.a.d.wijnands@uva.nl},
       E. Rol$^2$, E. Cackett$^3$, R.L.C. Starling$^2$, R.A. Remillard$^4$ \\ 
       $^1$
       Astronomical Institute ``Anton Pannekoek'', University of
       Amsterdam, Kruislaan 403, 1098 SJ, The Netherlands\\ 
       $^2$
       Department of Physics and Astronomy, University of Leicester,
       University Road, Leicester LE1 7RH, United Kingdom\\ 
       $^3$
       Department of Astronomy, University of Michigan, 500 Church St,
       Ann Arbor, MI 48109-1042, USA \\
       $^4$ MIT Kavli Center for Astrophysics and Space Research, Massachusetts
       Institute of Technology, Cambridge, MA 02139, USA}

\begin{document}



\maketitle

\label{firstpage}

\begin{abstract}
We present an analysis of the {\it Swift} BAT and XRT data of
GRB060602B, which is most likely an accreting neutron star in a binary
system and not a gamma-ray burst. Our analysis shows that the BAT
burst spectrum is consistent with a thermonuclear flash (type-I X-ray
burst) from the surface of an accreting neutron star in a binary
system. The X-ray binary nature is further confirmed by the report of
a detection of a faint point source at the position of the XRT
counterpart of the burst in archival {\it XMM-Newton} data
approximately 6 years before the burst and in more recent {\it
XMM-Newton} data obtained at the end of September 2006 (nearly 4
months after the burst). Since the source is very likely not a
gamma-ray burst, we rename the source Swift J1749.4--2807, based on
the {\it Swift}/BAT discovery coordinates. Using the BAT data of the
type-I X-ray burst we determined that the source is at most at a
distance of $6.7\pm1.3$ kpc. For a transiently accreting X-ray binary
its soft X-ray behaviour is atypical: its 2--10 keV X-ray luminosity
(as measured using the {\it Swift}/XRT data) decreased by nearly 3
orders of magnitude in about 1 day, much faster than what is usually
seen for X-ray transients. If the earlier phases of the outburst also
evolved this rapidly, then many similar systems might remain
undiscovered because the X-rays are difficult to detect and the type-I
X-ray bursts might be missed by all sky surveying instruments. This
source might be part of a class of very-fast transient low-mass X-ray
binary systems of which there may be a significant population in our
Galaxy.
\end{abstract}


\section{Introduction}

The primary goal of the {\it Swift} Gamma-Ray Burst mission
\citep{2004ApJ...611.1005G} is to discover and study gamma-ray bursts
(GRBs). Typically, GRBs are discovered with the Burst Alert Telescope
\citep[BAT;][]{2005SSRv..120..143B} and then the
satellite quickly slews toward the direction of the burst to
facilitate observations with the X-ray telescope
\citep[XRT;][]{2005SSRv..120..165B} and the UV/Optical telescope
\citep[UVOT;][]{2005SSRv..120...95R}. This allows detailed 
studies of the X-ray and UV/optical afterglows of the GRBs. In
addition to detecting GRBs, the BAT also detects persistent and
transient hard X-ray/soft gamma-ray sources and type-I X-ray bursts
from accreting neutron stars in low-mass X-ray binaries (LMXBs). Very
occasionally, when the BAT discovers a burst, it is not immediately
clear if it is a GRB or due to some other event such as a type-I X-ray
burst. The latter is the case of GRB060602B.

\subsection{GRB060602B}

On 2 June 2006, the {\it Swift} BAT detected a new burst named
GRB060602B \citep{2006GCN..5200....1S} at a position of R.A. =
17$^h$49$^m$28.2$^s$ and Dec. = --28$^\circ$07$'$15.5$''$ (J2000; with
a 90\% confidence error of 1.4$'$; \citealt{2006GCN..5208....1P}). The
burst lasted about 12 seconds and was strongest in the 15--25 keV
energy range and not seen above 50 keV
\citep{2006GCN..5208....1P}. The BAT spectrum could be fitted with a
power-law model with a photon index of $\sim$5 indicating a rather
soft spectrum for a GRB and suggesting that the source could be an
accreting neutron-star X-ray binary which exhibited a type-I X-ray
burst \citep[][also based on a Galactic position of $l$ = 1.15$^\circ$
and $b$ = -0.30$^\circ$]{2006GCN..5208....1P}. The short duration of
the burst is consistent with it being a type-I X-ray burst
\citep{2006astro.ph..8259G}. Just 83 seconds after
the BAT trigger, the {\it Swift} XRT and UVOT telescopes began taking
data of the source. No UVOT counterpart was seen but a decreasing
faint X-ray source was detected
\citep{2006GCN..5200....1S,2006GCN..5209....1B} which first slightly
increased in X-ray flux until 200 seconds after the trigger when it
then decreased in flux following a simple power-law decay with an
index of approximately $-1$ \citep{2006GCN..5209....1B}. The XRT
spectrum could be described by an absorbed power-law model with a
photon index of 3.1.  \citet{2006GCN..5209....1B} stated that the
spectral and temporal properties of the source are difficult to
reconcile with standard GRB afterglow models which would indicate that
the source might indeed be a Galactic accreting neutron star. This
conclusion was further strengthened by the detection of a faint X-ray
source at the XRT position in archival {\it XMM-Newton} data taken
nearly 6 years before the occurrence of the hard X-ray burst
\citep{2006GCN..5210....1H}.  The source was also officially retracted
as a GRB \citep{2007GCN..6013....1B}. To investigate the nature of
this source, we analysed in detail all available {\it Swift} BAT and
XRT data of the source as well as several archival {\it XMM-Newton}
observations.

\section{{\itshape Swift} data analysis and results}

\begin{figure} 
    \includegraphics[width=0.45\textwidth]{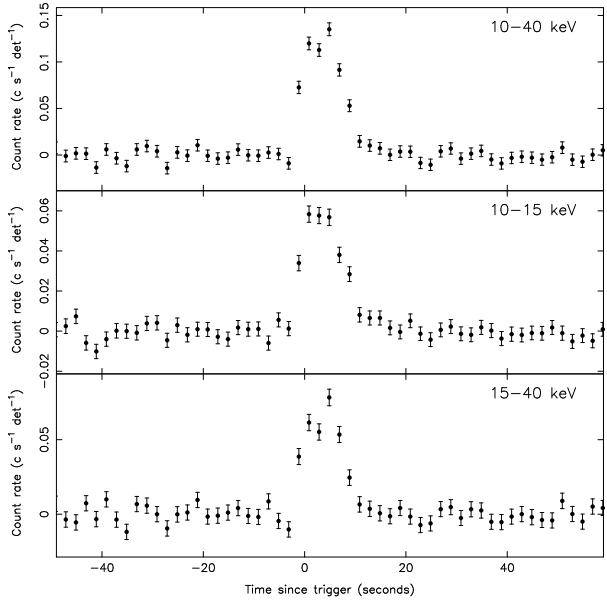}
      \caption{The {\it Swift}/BAT light curve of 
	GRB060602B in the
      energy range 10--40 keV (top panel), 10--15 keV (middle panel)
      and 15--40 keV (bottom panel).}  
     \label{fig:batlc} 
    \end{figure}

We analysed the BAT data of the burst using the standard
threads\footnote{See the data analysis documents for the {\it Swift}
instruments at http://swift.gsfc.nasa.gov/docs/swift/analysis/
\label{footnote}}. The burst light curve between 15 and 40 keV is 
shown in Figure~\ref{fig:batlc}. The burst could not be conclusively
detected above 40 keV indicating a very soft burst \citep[see
also][]{2006GCN..5208....1P}. Although the BAT is not calibrated below
15 keV, we also show the 10-15 keV and the 10-40 keV light curves to
demonstrate that the source had a rather high count rate at the lowest
energies despite the fact that the sensitivity of the BAT drops
significantly at these energies. This again points to a very soft
spectral shape of the burst. The burst lasted approximately 10 seconds
and it looks similar at different energies, although the statistics
are such that we cannot rule out the same spectral variability which
is normally observed for type-I X-ray bursts.  The burst profile does
not fully resemble the fast rise, exponential decay shape typically
seen in type-I X-ray bursts. However, this does not rule out a type-I
X-ray burst nature because our statistics are rather limited so
stringent conclusion about the burst profile cannot be made. In
addition type-I X-ray bursts at energies above 15 keV can have more
complex burst profiles (see, e.g., the type-I X-ray bursts seen from
4U 0614+09 seen by \citealt{2008ApJ...672L..37S} using BAT and the
hard X-ray bursts seen with INTEGRAL as reported by
\citealt{2006AstL...32..456C}).

\begin{figure}
   \centering
      \includegraphics[width=0.35\textwidth,angle=-90]{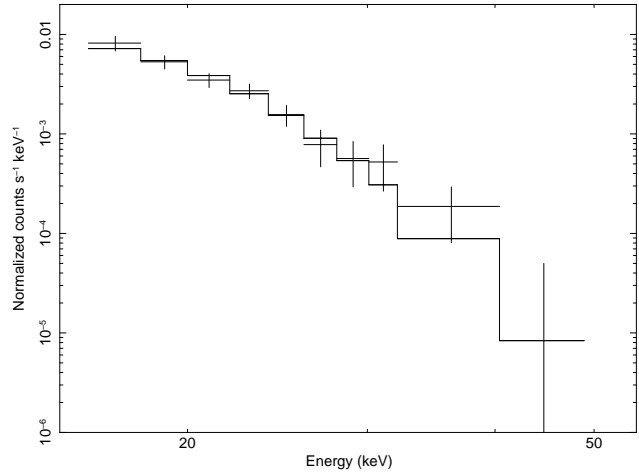}

      \caption{The {\it Swift}/BAT spectrum of GRB060602B. The points
      are the BAT data points and the solid line is the best fit
      black-body model through the data.} \label{fig:batspectrum}
\end{figure}

We extracted the spectrum of the whole burst and created the response
matrix as outlined in the threads. We fitted the resulting spectrum
(see Fig.~\ref{fig:batspectrum}) using Xspec between 15 and 50 keV. As
shown by \cite{2006GCN..5208....1P} the BAT spectrum could be fitted
well using a simple power-law model (with $\chi^2 = 14.2$ for 14
degrees of freedom [d.o.f.]). However, the power-law index obtained
was $\sim$5 which suggested a thermal spectral shape. Therefore, we
fitted the data with a black-body model and we obtained a temperature
of $2.9^{+0.4}_{-0.3}$ keV and a 15--50 keV flux of $1.7\pm0.1\times
10^{-8}$ erg s$^{-1}$ cm$^{-2}$ (with $\chi^2/d.o.f. =
11.9/14$). Extrapolating the flux to the energy range 0.01--100 keV
results in an approximate bolometric flux of $7^{+4}_{-2}\times
10^{-8}$ erg s$^{-1}$ cm$^{-2}$. The inferred radius of the black body
would be $8^{+4}_{-3}$ km for an assumed distance of 8 kpc.

\begin{figure*}
   \centering
      \includegraphics[width=\textwidth]{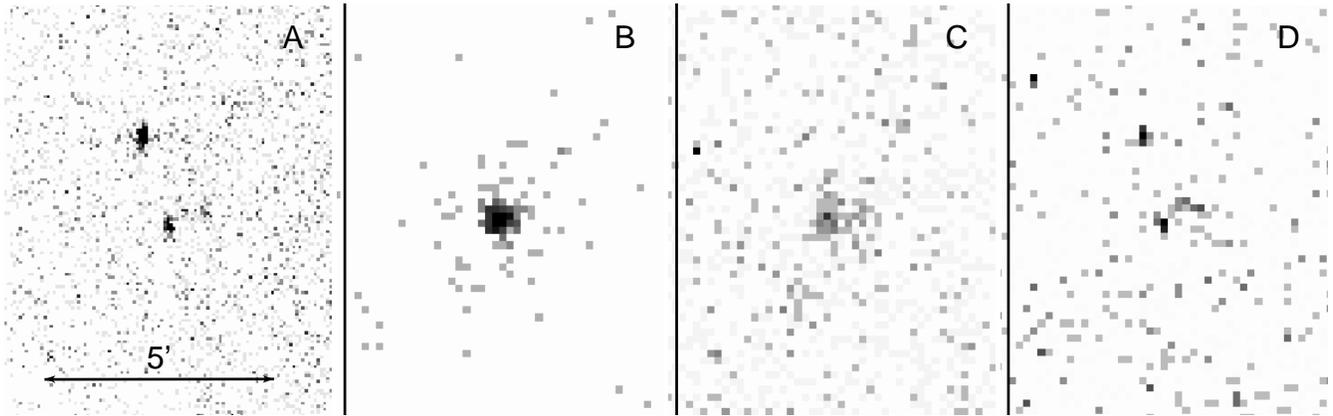}
      \caption{The {\it XMM-Newton} (left; A) and {\it Swift}/XRT
      (right three panels; B--D) images of GRB060602B. In panel B the 
      first $\sim$910 s of data taken on June 2 are
      shown (data set 1 and 2; ObsId 002131900), in panel C the
      remaining data of June 2 (data set 3), and in D the combined
      data of June 4 to June 10 (ObsID
      00213190001-00213190006). } \label{fig:images}
\end{figure*}

About 83 seconds after the burst was first detected with the BAT, the
XRT started to observe the source \citep{2006GCN..5200....1S}. During
the next 8 days a total of $\sim$55 ksec of data was obtained from
this field. A log of the observations is shown in
Table~\ref{tab:Swift-log}. During all observations the photon counting
mode was used. All the XRT observations were reprocessed using the
standard method (see footnote~\ref{footnote}). Each observation was
subdivided into multiple data segments (ranging from a few hundred
seconds in length to about a ksec). During the first observation a
relatively bright source was detected which decayed rapidly. During
the remaining observations the source was very faint but could still
be detected when combining multiple observations (see
Fig.~\ref{fig:images}).  We found that below 1.5 keV hardly any source
counts were present (likely due to the relatively high Galactic
absorption) so we only used data between 1.5 and 7 keV to limit the
effects of the background on the statistics (above 7 keV, the data
were dominated by the background).  The coordinates of the source as
obtained from the first observation are R.A. = 17$^h$49$^m$31.89$^s$
and Dec. = --28$^\circ$08$'$02.8$''$ (J2000; with a 90\% confidence
error of 3.7$''$).  This position is consistent with the revised
position of R.A. = 17$^h$49$^m$31.94$^s$ and Dec. =
--28$^\circ$08$'$05.8$''$ (J2000; with a 90\% confidence error of
3.3$''$;) reported by \cite{2007AJ....133.1027B}. We extracted the
light curve for the source using a variable extraction region
(dependent on the brightness of the source to optimise the
signal-to-noise) and background subtracted the count rates. We
adaptively binned the data so as to have 20 counts per bin (grouping
was done separately for the data segments of observation 00213190000
because of the rapid decrease in count rate).

\begin{figure}
   \centering
     \includegraphics[width=0.35\textwidth,angle=-90]{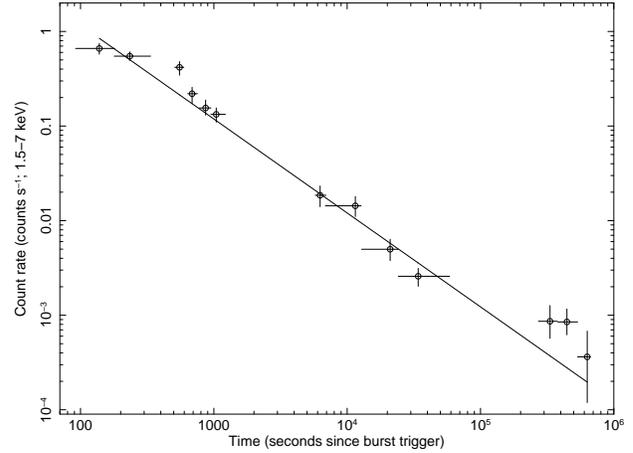}

      \caption{The {\it Swift}/XRT light curve of GRB060602B (1.5--7
      keV). The solid line is the best power-law decay model.}
      \label{fig:xrtlc}
   \end{figure}

Initially we found that the count rate slightly increased during the
first $\sim$200 seconds, decreasing steadily after that
\citep[see also][]{2006GCN..5200....1S}. However, when we checked 
the source profile obtained and compared it with the expected profile
using the point-spread-function of the XRT, we found that the source
likely suffered from pile-up during the initial part of the light
curve. We estimated the amount of pile-up by comparing the two
profiles and we corrected the count rates for it. The resulting light
curves are shown in Figure~\ref{fig:xrtlc}. Clearly, the initial rise
has disappeared and the source decreases in flux from the start of the
XRT observations until it reached a more constant level in the later
observations. We fitted the resulting decay curve using different
models and found that a simple power-law decay model with an index of
$-0.99 \pm 0.05$ fitted the decay curve best
\citep[Fig.~\ref{fig:xrtlc}; see also][]{2006GCN..5209....1B},
although formally still not acceptable ($\chi^2 = 28.4$, $d.o.f. =
11$). This is due to the third point and the last few points in the
decay light curve which are significantly above the general decay
trend. This indicates that the decay was not perfectly smooth and that
possible small flares occurred on top of the power-law decay.

Adding a constant level at the end of the decay did not significantly
improve the fit significantly (the resulting $\chi^2= 26.1$ for 10
$d.o.f.$); the index obtained was again $\sim -1$. We also tried
fitting an exponential decay function (with and without a levelling
off at the end; $\chi^2/d.o.f =83.6/10$ and $\chi^2/d.o.f. = 52.2/9$,
respectively), but no such model could reproduce the data.

The investigation of the X-ray spectrum of the source was complicated
by 3 factors: the fact that the source was rapidly decaying (which
might possibly be accompanied by spectral changes), the pile-up during
the first $\sim$1000 seconds of the data, and the very faint fluxes in
the late stages of the decay. We focused on the data of observation
00213190000 and divided it into three data sets: the first set
contained the first $\sim$250 seconds of data of this observation, the
second set contained the next $\sim$660 seconds of data (starting
about 550 seconds after the beginning of the observation), and the
third set contained the remaining data ($\sim$16 ksec of exposure time
spread out over 53 ksec). The first data set was most affected by
pile-up so we extracted the spectrum using only GRADES 0 and using an
annulus extraction region with inner radius of 8 pixels and outer
radius of 30 pixels. The second data set suffered from less severe
pile-up, so the annulus extraction region had an inner radius of 3
pixels (and the same outer radius; also only GRADES 0 were
extracted). For the third set the pile-up was negligible and we used a
circular extraction region with a radius of 20 pixels and GRADES 0 to
12. We used the pre-made response matrices but we created our own
ancillary response matrices which take into account the size and shape
of the extraction region. For the background spectrum we used four
source-free circular extraction regions, all with radii of 20 pixels,
and combined the data from all four regions.  After rebinning the
spectra to have 10 counts per bin, we fitted them using Xspec.  The
observations 00213190001 to 00213190006 could not be used to
investigate the spectrum because, even though the source was detected
when combining these observations (Fig.~\ref{fig:images}), the number
of counts detected were not enough to obtain a useful spectrum.

\begin{figure}
   \centering
      \includegraphics[width=0.35\textwidth,angle=-90]{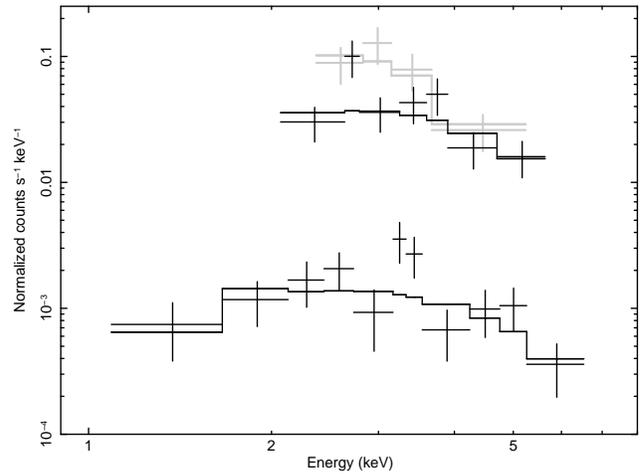}
      \caption{The {\it Swift}/XRT spectrum of GRB060602B during
      observation 00213190000: the top two data points are for data
      set 1 (grey) and 2 (black); the bottom data set is set 3. The
      solid lines through the data points represent the best fit model.}
	 \label{fig:xrtspectrum}
\end{figure}

The three individual data sets from observation 00213190000 could be
well fitted with a single component model (e.g., a disk black-body or
power-law model; reduced $\chi^2$ around 0.8-1.0), but when fitting
the first data set with an absorbed power-law model the resulting
index was $>$6 and a column density $N_{\rm H} > 10^{23}$ cm$^{-2}$
(albeit with large errors). These values indicate that the spectrum
most likely has a thermal shape instead of a non-thermal one (although
we note that a thermal model and a power-law model fit the data
equally well). The other two data sets were consistent with a
power-law model with indices between 2 and 3. To obtain the best
constraints on the fit parameters, we fitted the three data sets
simultaneously and tied the column density $N_{\rm H}$ between the
observations. For the first data set we used a disk black-body model
but for the other two a power-law model. This combination of models
fitted the data reasonably well ($\chi^2/d.o.f. = 19.2/15$) and the
resulting fit parameters are listed in Table~\ref{tab:Swift-results}
(see also Fig.~\ref{fig:xrtspectrum}). We used a disk black body model
in the first, brightest part of the observation and a power-law model
for the fainter parts because such models are commonly used for X-ray
transients at similar brightness levels.  We note however, that the
obtained parameters using these models are just an indication of the
spectral shape of the source since many different types of models or
combination of models fit the data equally well. The only tentative
conclusion we can draw from these data is that the source seemed to
switch from a thermal-like spectrum to a non-thermal spectrum during
the first hundreds of seconds of the decay. Because of the large
uncertainties in the fit parameter and in determining which model
should be used to fit the data, the errors on the absorbed fluxes are
very large (several tens of percent) and even larger on the unabsorbed
fluxes. The fluxes quoted in Table~\ref{tab:Swift-results} are not
corrected for absorption; typically correction factors range between
1.3 and 1.6. Due to uncertainties of extrapolating the model outside
the fitting range, we restrict ourselves to quoting only the 2--10 keV
fluxes.

\section{{\itshape XMM-Newton} data analysis and results}

The position of GRB060602B was in the field-of-view of three {\it
XMM-Newton}
\citep{2001A&A...365L...1J,2001A&A...365L..27T,2001A&A...365L..18S}
observations (see Tab.~\ref{tab:XMM-log}).  One observation was before
the burst detected from the source (nearly six years before;
Fig.~\ref{fig:images}; see also
\citealt{2006GCN..5210....1H}) and the source is listed in the
Second {\it XMM-Newton} Serendipitous Source
Catalogue\footnote{http://xmmssc-www.star.le.ac.uk/Catalogue/2XMM/}
(\citealt{watson}; provided by the {\it XMM-Newton} Survey Science
Centre) as 2XMM J174931.6--280805.  The other two were performed
almost four months after the {\it Swift} observations. We analysed
these observations using SAS\footnote{http://xmm.vilspa.esa.es/sas/;
version 7.0.0}. All instruments were active but here we only discuss
the data as obtained with the European Photon Imaging Camera (EPIC)
instruments (due to the very low flux of the source, it was not
detected in the RGS instruments). For all data sets the MOS cameras
were operating in full window mode. In contrast, the pn camera was
used in full window mode only for September 23, 2000; for the other
two dates it was in timing mode which is not well suited to study
faint sources. Therefore, we only used the pn data obtained during the
first observation.

We searched for background flares using light curves for photon
energies above 10 keV. During the first observation, one bright flare
was present and we removed it from the data (this removed several
hundreds of seconds of data). During the other two observations no
background flares were present and all data could be used. For each
observation, we combined the data of the EPIC cameras to obtain the
highest signal-to-noise ratio for the determination of the source
position.  Due to the high offset angle and the faintness of the
source, the {\it XMM-Newton} positions have a relatively large error
(up to 4$``$) and the positions are not better than the one we
obtained using {\it Swift} when the source was in outburst. However,
during all three observations, the position of the faint {\it
XMM-Newton} source is consistent with the {\it Swift}/XRT position of
GRB060602B. The count rates of the source were extracted using the
funcnts program from the funtools package. The count rates were
background subtracted and exposure corrected. As source extraction
region we used a circle with a radius of 15$''$ and as background
region a circle with a radius of 2$'$ in a region free of sources
close to the position of GRB060602B. The observed count rates are
listed in Table~\ref{tab:XMM-results}. The low count rates observed
for the source did not allow for a spectral analysis. We estimated the
source flux using WebPIMMS\footnote{Available at
http://heasarc.gsfc.nasa.gov/Tools/w3pimms.html} and assuming an
absorbed power-law spectral shape with a column density of $3\times
10^{22}$ cm$^{-2}$ (as measured for the {\it Swift}/XRT outburst data;
see Tab.~\ref{tab:Swift-results}) and a power-law index of 2. The
resulting fluxes are also listed in Table~\ref{tab:XMM-results}. We
also calculated the expected {\it Swift}/XRT count rate for the last
two observations; the obtained 1.5--7 keV {\it Swift}/XRT count rates
are $1-2 \times 10^{-3}$ count s$^{-1}$ which is a bit higher than
observed at the end of the outburst with the XRT but is likely
consistent when taking into account the many uncertainties and
assumptions made. Therefore, {\it Swift} very likely observed the full
decay outburst of this source, all the way down to quiescence.

\section{Non-detection of the source by the all-sky monitor aboard 
{\itshape RXTE} }

The all-sky-monitor (ASM; \citealt{1996ApJ...469L..33L}; which is
aboard the {\itshape Rossi X-ray Timing Explorer} [{\itshape RXTE};
\citealt{1993A&AS...97..355B}]) light curve for GRB060602B was
determined over the duration of the {\it RXTE} mission (1996 January
to 2007 August).  The systematic uncertainties are important in this
case, since the source is only 1.2$^\circ$ from the Galactic centre,
and 0.22$^\circ$ from the X-ray transient IGR J17497--2821
\citep{2007A&A...461L..17W}\footnote{IGR J17497--2821 was in outburst
in September/October 2006 and we detected this outburst with the ASM
and observed a maximum flux of 25 mCrab (1.5--12 keV; 1 week bins),
which is near the threshold for the ASM sensitivity at the Galactic
centre}. In this region of the sky, an average of 24 X-ray sources
must be included in the coded mask deconvolution for each 90 s
exposure by one of the ASM cameras.  For GRB060602B, the ASM light
curve shows no significant detections in either 90 s exposures or in
weekly intervals.  In particular, on 2006 June 2 there are only 5 ASM
camera exposures with an average flux of 34$\pm$18 mCrab.  The weekly
exposures during 2006 May and June are near average levels, with upper
limits in the range 20-35 mCrab (1.5--12 keV) per week over this
interval.

\section{Discussion}

We reported on the {\it Swift} BAT and XRT data of the X-ray burst
source GRB060602B. The BAT spectrum could well be fitted with a
black-body model with a temperature of $\sim$2.9 keV and inferred
radius of $\sim$8 km, which is in the range of temperatures and radii
seen from type-I X-ray bursts from neutron stars in LMXBs \citep[see,
e.g., ][ and references
therein]{1993SSRv...62..223L,2003A&A...399..663K}. This result,
together with the detections of the source (with a roughly constant
luminosity) in archival (six years before the burst) and more recently
obtained (four months after) {\it XMM-Newton} data strongly indicate
that the source is indeed not a gamma-ray burst but an accreting
neutron star (as first suggested by \citealt{2006GCN..5208....1P} and
\citealt{2006GCN..5210....1H}). We rename the source Swift
J1749.4--2807 based on the revised BAT discovery coordinates
\citep{2006GCN..5208....1P} and we will use this name from now on.

The nature of the donor star in this system is unclear. If the donor
is a relatively high mass star ($>10 M_\odot$), Swift J1749.4--2807
might be a member of the group of recently identified supergiant fast
X-ray transients \cite[see, e.g.,][and references
therein]{2006ApJ...646..452S}. These are systems harbouring likely a
OB supergiant and they exhibit fast X-ray outbursts. However, despite
that the typical outburst luminosities of these systems is around
$10^{36}$ erg s$^{-1}$, which is similar to what we see for Swift
J1749.4--2807, the duration of the outbursts of the supergiant fast
X-ray transients is generally much shorter \cite[only a few minutes to
at most a few hours; e.g.,][although some outbursts lasted
significantly longer but the majority have a very short
duration]{2006ApJ...646..452S} and the outbursts are more erratic than
what we have observed for Swift J1749.4--2807. Moreover, no high-mass
X-ray binary has so far ever exhibited a type-I X-ray burst. Although
burst-like events have been observed from for example SMC X-1, such
events have typical X-ray spectra which are not consistent with a
black-body model \citep{1991ApJ...371..332A}.  Therefore, we consider
it most likely that this system has a low-mass companion star (with
mass $<1 M_\odot$) and that Swift J1749.4--2807 is not a supergiant
fast X-ray transient.

If the BAT burst was indeed a type-I X-ray burst, we can obtain a
distance estimate towards the source. Assuming that the Eddington
limit was reached during the burst and that the burst ignited in a
hydrogen-poor environment, we can use the empirically determined
Eddington luminosity by \citet[$3.8\times10^{38}$ erg
s$^{-1}$]{2003A&A...399..663K} to obtain a distance of $6.7\pm 1.3$
kpc. If we instead use equation 6 of \citet{2006astro.ph..8259G} we
obtain a distance of $5.6\pm1.1$ kpc for hydrogen-poor bursts and
$4.3\pm0.9$ kpc for hydrogen-rich bursts (assuming a neutron star with
a mass of 1.4 $M_\odot$ and a radius of 10 km). We note that the
fluxes we obtain are those averaged over the whole burst and therefore
it is possible that the peak flux was even higher and thus the
distance smaller. Furthermore, if the burst we observed was not
Eddington limited, then the distance would also be smaller.

Although the BAT burst spectrum strongly suggests that the source is a
Galactic accreting neutron star, the XRT data are atypical for what is
observed for ordinary neutron-star X-ray transients. Such systems are
typically active (when in outburst) for weeks to months (some even
years to decades) with a decay time scale (i.e., the e-folding time)
of at least a few days to weeks \citep[see
e.g.][]{1997ApJ...491..312C,1998ApJ...499L..65C,2003MNRAS.341..823J}. In
contrast, within $\sim$1 day, Swift J1749.4--2807 decreased in flux by
three orders of magnitude: from close to $10^{-10}$ erg s$^{-1}$
cm$^{-2}$ (the 2--10 keV flux during the first XRT data set) to about
$10^{-13}$ erg s$^{-1}$ cm$^{-2}$ (as measured during the last XRT
data and the archival {\it XMM-Newton} observations). Assuming a
distance of 6.7 kpc, the measured peak luminosity was close to
$5\times 10^{35}$ erg s$^{-1}$ (for the energy range 2--10
keV). However, if the X-ray flux began already to decay at a similar
rate at the end of the type-I X-ray burst, then the source must have
been accreting at about 10 times higher luminosity ($\sim5\times
10^{36}$ erg s$^{-1}$) when the burst occurred. The latter is a
typical X-ray luminosity for the known active bursting sources in our
Galaxy. Note that if the decay started at the end of the burst, then
the source decreased by nearly four orders of magnitude in flux in one
day!  However, with the current data it cannot be assessed whether or
not the source had a similar decay rate in the time between the type-I
X-ray burst and the start of the XRT observations. Furthermore, it is
also unclear if the occurrence of the burst in some way triggered the
decay of the source or if the two are unrelated. The observed
tentative change in spectral shape has been seen for many neutron-star
X-ray transients. Those systems typically change their spectral shape
from thermally dominated to non-thermal dominated around a few times
$10^{36}$ erg s$^{-1}$
\citep[see e.g.][]{2003MNRAS.338..189M,2007MNRAS.378...13G}, which is
close to the X-ray luminosity at which we observe the spectral shape
of Swift J1749.4--2807 to change.

The exact duration of the outburst is also unclear. It is possible
that the source was active for a considerable amount of time (days to
even weeks) before the BAT burst occurred. If true, the 2--10 keV flux
of the source before the burst should not have been much above
approximately $5 \times 10^{-10}$ erg s$^{-1}$ (corresponding to a few
times $10^{36}$ erg s$^{-1}$) otherwise we would have detected the
source with the {\it RXTE}/ASM. It is also possible that the rise and
the peak were as fast as the decay observed in this source. For a
type-I X-ray burst to occur, a certain amount of matter must be
accreted. However, for ordinary neutron star LMXBs the bursts can
recur within hours to a day when they have luminosities similar to
those observed for our source when the burst occurred
\citep{2006astro.ph..8259G}. Therefore, it is possible that the source
was active for only a day or so and still accumulated enough matter to
exhibit the burst and then disappeared again. Sources which exhibit
such short outbursts are easily missed by monitoring instruments. This
could then indicate that a significant number of similar systems may
be present in our Galaxy but which are usually missed when they are in
outburst. Their faint accretion luminosity and their short outbursts
might make them very difficult to detect with monitoring instruments;
their bright type-I X-ray bursts might also easily be
missed.\footnote{We can also speculate that a class of similarly very
fast transients are present in the Galaxy which harbour a black hole
instead of a neutron star. Such systems would be even more difficult
to find because the discovery characteristic, the type-I X-ray bursts,
which allowed the known systems to be found, do not occur for black
hole systems.}

Interestingly, there is one class of neutron-star X-ray binaries which
might be such a class of sources and which might be related to Swift
J1749.4--2807: the so-called burst-only sources \cite[see][for an
overview of these sources]{2004NuPhS.132..518C}. These systems are
accreting neutron star sources which were discovered (mostly with {\it
BeppoSAX} but also with {\it INTEGRAL}) because a type-I X-ray burst
was detected from them but which could not be detected outside the
bursts with any of the monitoring instruments in orbit. The accretion
luminosities of these sources at the time of the type-I X-ray bursts
should be below $\sim$$10^{36}$ erg s$^{-1}$ for them to remain
undetectable. More sensitive follow-up observations with for example
{\it Chandra} or {\it XMM-Newton} found that although some are
persistent sources with very low luminosities, most of them were
likely neutron-star transients which most of the time were in a very
dim quiescent state (with X-ray luminosities of the order of $10^{32}$
erg s$^{-1}$ or less; see
\citealt{2002A&A...392..931C,2004NuPhS.132..518C}). In such systems, 
the type-I bursts were seen during one of their very-faint X-ray
outbursts.

One of these burst-only sources (called SAX J2224.9+5421;
\citealt{2002A&A...392..885C}) was of particular interest because within
8 hours after the Wide Field Camera of {\it BeppoSAX} discovered it
through its burst\footnote{We note that the type-I X-ray burst
nature of this {\it BeppoSAX} burst could not conclusively be
established \citep{2002A&A...392..885C} and it is possible that
the source is of a different, as yet unknown, origin. However, for the
current paper we assume it was indeed a type-I X-ray burst.}, {\it
BeppoSAX} pointed at the source using the Narrow Field Instrument
(NFI) and could detect the source only at a 2--10 keV flux of $\sim
1.3\times 10^{-13}$ erg s$^{-1}$ cm$^{-2}$
\citep{1999GCN...445....1A} resulting in a luminosity of $\sim 8
\times 10^{32}$ erg s$^{-1}$ 
\citep[assuming a distance of 7.1 kpc;][]{2002A&A...392..885C}.
This detection of the source at a very faint level only 8 hours after
the occurrence of the burst is very reminiscent of what we have
observed for Swift J1749.4--2807. Only $\sim$8 hours after the burst,
the XRT 2--10 keV flux of Swift J1749.4--2807 had decreased already to
around $5\times 10^{-13}$ erg s$^{-1}$ cm$^{-2}$, which is of the same
order of magnitude as the flux observed for SAX J2224.9+5421 outside
its burst. \citet{2002A&A...392..931C} suggested that SAX J2224.9+5421
could be bursting at very low (near quiescent) X-ray luminosities, but
our results on Swift J1749.4--2807 also suggest that both sources
could be very similar sources which exhibit a relatively faint (but
not very faint) outburst but they decay very rapidly after the
occurrence of their type-I X-ray bursts.

Determining the exact accretion luminosity at which the bursts occur
is important in understanding the burst physics since the burst
properties depend strongly on the accretion rate at the time the burst
occurs. Although the accretion situation is evident for Swift
J1749.4--2807, it remains unclear for SAX J2224.9+5421. Clearly, for
these systems and others similar to them, the very short slew time
available with {\it Swift} is necessary to distinguish between the
different scenarios. 

Swift J1749.4--2807 decayed rapidly to a constant level which was very
similar to what {\it XMM-Newton} saw from the source six years before
the occurrence of the burst and four months after it. Therefore, this
constant flux level very likely represents the quiescent flux of the
source which, for a distance of 6.7 kpc, results in a 2--10 keV
luminosity of 0.5--1.0 $\times 10^{33}$ erg s$^{-1}$
\citep[see also][]{2006GCN..5210....1H}. This is very similar to the
quiescent luminosity seen for other neutron-star X-ray transients in
their quiescent state. Sadly, due to the faintness of the source no
spectral information could be obtained but the high $N_{\rm H}$ (as
measured in outburst with the XRT) in-front of the source makes it
difficult to detect any soft, thermal component and it is very likely
that the emission we observe is (mostly) due to a non-thermal
component. With the current data no sensible upper limits can be
obtained on any thermal component with which we could test cooling
models for accretion-heated neutron stars. A longer exposure
observation with {\it XMM-Newton} (with the source on-axis) or a deep
{\it Chandra} observation (with its much lower background) is needed
to study the quiescent emission of this source with the detail
necessary to allow comparative studies with other quiescent
neutron-star X-ray transients.

\section*{Acknowledgements}

We acknowledge the use of public data from the {\it Swift} data
archive. RCLS acknowledge financial support from PPARC.

\newpage\clearpage

\begin{table}
	\caption{The log of the {\it Swift} observations\label{tab:Swift-log}}
	\begin{tabular}{l|c|l}
	ObsId & Date (June 2006) & Detectors used in analysis\\
	\hline
        00213190000 &  02  at 23:39 & XRT, BAT \\ 
	00213190001 &  04  at 00:18 & XRT \\
	00213190002 &  06  at 03:03 & XRT \\
	00213190003 &  07  at 03:09 & XRT \\
	00213190004 &  08  at 03:15 & XRT \\
	00213190005 &  09  at 00:22 & XRT \\
	00213190006 &  10  at 00:25 & XRT \\
	\hline
	\end{tabular}
\end{table}

\begin{table}
	\caption{The log of the {\it XMM-Newton} observations\label{tab:XMM-log}}
	\begin{tabular}{lcll}
	ObsId & Date & Detectors used & Filter\\
	\hline
	0112980101 & 23 September 2000 & MOS1, MOS2, pn & Medium\\
        0410580401 & 22 September 2006 & MOS1, MOS2     & Thick\\
        0410580501 & 26 September 2006 & MOS1, MOS2     & Thick\\
	\end{tabular}
\end{table}

\newpage\clearpage

\begin{table}
	\caption{The results of the {\it Swift}/XRT spectral analysis\label{tab:Swift-results}}
	\begin{tabular}{lcccc}
	Set & $N_{\rm H}$$^{1}$            &  $kT$               & $\Gamma$            & Flux$^2$                 \\
	        &( $\times 10^{22}$ cm$^{-2}$) & (keV)               &                     & (erg s$^{-1}$ cm$^{-2}$) \\
	\hline
	1       &  3$^{+4}_{-2}$               & 0.8$^{+0.5}_{-0.2}$ &                     & $8.0\times 10^{-11}$     \\
        2       &                              &                     & 1.9$^{+1.5}_{-1.0}$ & $3.1\times 10^{-11}$     \\
        3       &                              &                     & 2.2$^{+1.9}_{-0.7}$ & $4.6\times 10^{-13}$     \\
	\end{tabular}
	$^1$ The column density was tied between the three data sets\\
	$^2$ For 2--10 keV and not corrected for absorption
\end{table}

\begin{table}
	\caption{The results of the {\it XMM-Newton} analysis\label{tab:XMM-results}}
	\begin{tabular}{lccc}
         Detector         &  0112980101 & 0410580401 & 0410580501\\
	\hline
        MOS1                                                   &             & & \\
          -- count rate ($\times 10^{-3}$ counts s$^{-1}$) & 4.4$\pm$1.1 & 2.5$\pm$0.7 & 3.4$\pm$0.8  \\
          -- absorbed flux                                     & 1.3$\pm$0.3 & 0.8$\pm$0.2 & 1.1$\pm$0.2   \\
	  -- unabsorbed flux                                   & 1.7$\pm$0.4 & 1.0$\pm$0.3 & 1.4$\pm$0.2  \\
        MOS2                                                   &             & & \\
          -- count rate ($\times 10^{-3}$ counts s$^{-1}$) & 4.9$\pm$1.1 & 5.0$\pm$1.0 & 3.5$\pm$0.9\\
          -- absorbed flux                                     & 1.5$\pm$0.3 & 1.5$\pm$0.4 & 1.1$\pm$0.3\\
	  -- unabsorbed flux                                   & 1.9$\pm$0.4 & 2.0$\pm$0.4 & 1.4$\pm$0.4\\
	pn                                                     &             & & \\
          -- count rate ($\times 10^{-3}$ counts s$^{-1}$) & 14$\pm$2    & & \\
          -- absorbed flux                                     & 1.4$\pm$0.2 &             &              \\
	  -- unabsorbed flux                                   & 1.8$\pm$0.3 &             &              \\
	\hline
	\end{tabular}
	Note: The count rates are for 0.2--12 keV for the MOS instruments and 0.3--12 keV for the pn detectors and
	the fluxes are for 2--10 keV and in units of $10^{-13}$ erg s$^{-1}$ cm$^{-2}$ and where calculated using WebPIMMS
        and an absorbed power-law model using $N_{\rm H} = 3\times 10^{22}$ cm$^{-2}$ and a power-law index of 2 
\end{table}

\end{document}